%Paper: gr-qc/9210009
%From: rah@thp.Uni-Koeln.DE (Ralf Hecht)
%Date: Thu, 15 Oct 92 12:52:48 MET
%Date (revised): Thu, 15 Oct 92 18:35:59 MET

%========================================================================
\magnification=\magstep1\hsize=13cm\vsize=20cm\overfullrule 0pt
\baselineskip=13pt plus1pt minus1pt
\lineskip=3.5pt plus1pt minus1pt
\lineskiplimit=3.5pt
\parskip=4pt plus1pt minus4pt

\def\vta{\vartheta}\def\a{\alpha}\def\b{\beta}\def\c{\gamma}
\def\d{\delta}

% macro for slash
\def\negenspace{\kern-1.1em}\def\quer{\negenspace\nearrow}

%\def \Behauptung{
%= \hbox to 0pt{ \kern -10pt \lower 3pt \vbox to 15pt {\hbox
%{$\scriptstyle ! \> {} $} \vglue 6pt \hbox {} }} }

%Macro for leftitems

\def\ltextindent#1{\hbox to \hangindent{#1\hss}\ignorespaces}

%Macro for section, subsection and equation numbers:

%Macro for d'Alembertian:

\def\sqr#1#2{{\vcenter{\hrule height.#2pt\hbox{\vrule width.#2pt
height#1pt \kern#1pt \vrule width.#2pt}\hrule height.#2pt}}}
\def\square{\mathchoice\sqr64\sqr64\sqr{4.2}3\sqr{3.0}3}

%Macro for footnotes:
\newcount\refno
\refno=1
\def\y{\the\refno}
\def\myfoot#1{\footnote{$^{(\y)}$}{#1}
                 \advance\refno by 1}

%Macro for list of references:

\def\newref{\vskip 1pc %\medskip
            \hangindent=2pc
            \hangafter=1
            \noindent}

%Macro for not equal:
\def\neq{\hbox{$\,$=\kern-6.5pt /$\,$}}

%Macro for = with * on top:

%Macro for semidirect product
\def\semidirect{\;{\rlap{$\subset$}\times}\;}
%Macro for boldface omega (\clom):
%AM font:
%\font\fbg=ambi10\def\clom{\hbox{\fbg\char33}}

%CM font (when using IPS):

%Macro for section, and equation numbers: (Use \sectio)

\newcount\secno
\secno=0
\newcount\fmno\def\z{\global\advance\fmno by 1 \the\secno.
                       \the\fmno}
\def\sectio#1{\medbreak\global\advance\secno by 1
                  \fmno=0
     \noindent{\the\secno. {\it #1}}\noindent}
%Pound sterling:= {\it \$\/}50--00.

%=====================================================================
\noindent{\it file con7.tex, 1992-10-14 }
\bigskip
\centerline{\bf IMPROVED ENERGY-MOMENTUM CURRENTS}
 \centerline{\bf IN METRIC-AFFINE SPACETIME}
\bigskip
\centerline{by}
\bigskip
\centerline{Ralf Hecht$^{*}$, Friedrich W. Hehl$^{*\$}$,
J. Dermott McCrea$^{**}$,}
\centerline{Eckehard W. Mielke$^{*\$ }$, and
Yuval Ne'eman$^{*** \diamond \$ +)}$}
\bigskip
\noindent $^{*})$ Institute for Theoretical Physics, University of
Cologne, D(W)-5000 K\"oln 41, Germany

\noindent $^{**})$ Department of Mathematical Physics,
University College, Dublin 4, and Dublin Institute for
Advanced Studies, Dublin 4, Ireland

\noindent $^{***})$ Institut des Hautes Etudes Scientifiques,
F-91400 Bures-sur-Yvette, France, and
Raymond and Beverley Sackler Faculty of
Exact Sciences, Tel-Aviv University, Tel-Aviv 69978, Israel
%\bigskip
\bigskip
\centerline{\bf Abstract}
\bigskip
In Minkowski spacetime it is well-known that the canonical
energy-momentum current is involved in the construction of the
globally conserved currents of energy-momentum and total angular
momentum. For the construction of conserved currents corresponding to
(approximate) scale and proper conformal symmetries, however, an
improved energy-momentum current is needed.  By extending the
Minkowskian framework to a genuine metric-affine spacetime, we find
that the affine Noether identities and the conformal Killing equations
enforce this improvement in a rather natural way. So far, no
gravitational dynamics is involved in our construction. The resulting
dilation and proper conformal currents are conserved provided the
trace of the energy-momentum current satisfies a (mild) scaling
relation or even vanishes.
\medskip
%\noindent PACS numbers: 02.40.+m, 04.50.+h, 12.25.+e .
\vfill
\noindent $^\diamond)$ On leave from the Center for Particle Physics,
University of Texas, Austin, Texas 78712, USA

\medskip
\noindent $^{\$})$ Supported by the German-Israeli Foundation for
Scientific Research \& Development (GIF), Jerusalem and Munich.

\noindent $^{+})$Supported in part by DOE Grant DE-FG05-85-ER40200.

\eject

\sectio{\bf Introduction}

If a spacetime admits symmetries, we can construct a
set of {\it invariant\/} conserved quantities, one for each symmetry.
Consider the Riemannian spacetime of general relativity (GR).
By a {\it Killing symmetry} we understand a diffeomorphism of the Riemannian
spacetime
under which the metric is invariant.  Let the
vector field $\xi=\xi^{\alpha}\, e_{\alpha}$ be the generator of such
a 1-parameter group of diffeomorphisms of the spacetime.
Then it obeys the Killing
equation\myfoot{Here $e_\a$ is an arbitrary vector basis of the
tangent space and $g_{\a\b}$ are the components of the metric. The Lie
derivative with respect to $\xi$ is denoted by ${\cal L}_\xi$, the
covariant exterior derivative by $D$, and the interior product by
$\rfloor$.}
$${\cal L}_\xi\, g=0\qquad\Rightarrow\qquad e_{(\a}\rfloor
{\buildrel {o}\over {D}}\xi_{\b)}=0\, , \eqno(\z)$$
with the Riemannian exterior covariant derivative
${\buildrel {o}\over {D}}$.

Following the standard procedure of GR,
see Penrose and Rindler [1], it is straightforward
to construct an energy-momentum current which is a closed form,
provided the matter field equation is satisfied.  Let $t_{\a\b}$ be the
symmetric energy-mo\-men\-tum tensor of matter and $t_\a$ the
corresponding covariantly conserved energy-momentum 3-form\myfoot{If
$\vta^\a$ is the 1-form basis dual to $e_\a$, we have
$t_\a=t_{\a\b}\,\eta^\b $ where $\eta^\a ={}^\ast\vta^\a$ and
$\ast$ denotes the Hodge star. Moreover, $\eta^{\a\b}={}^\ast(\vta^\a
\wedge\vta^\b)$.}. Then the energy current
$\varepsilon_ {\rm \, R}$ represents a weakly closed 3-form:
$$\varepsilon_{\rm \, R}:=\xi^\a\, t_\a\,,\qquad\qquad
d\,\varepsilon_{\rm \, R}\cong 0\,.\eqno(\z)$$
{\it Weakly} means that the
last equation is only valid, if the matter field equation
holds. Accordingly, for a timelike Killing vector field,
$\varepsilon_{\rm \, R}$ yields, after integration over a spacelike
hypersurface $S$, the conserved energy $\int_{S}\varepsilon_{\rm \, R}$.

For a bosonic description of gravitational interactions, the
metric-affine gauge theory of gravity (MAG) is a very general framework,
see Ref.[2]. The spacetime continuum of MAG is represented by a
differentiable manifold $(L_4 , g)$ with a coframe $\vta^\alpha$, an
arbitrary linear connection $\Gamma_\alpha{}^\beta$, and a second rank
symmetric tensor field, the metric $g=g_{\alpha\beta}\,\vta^\alpha\otimes
\vta^\beta$. In extending the previous notion, we will now understand
by a Killing symmetry a diffeomorphism under
which the metric {\it and} the connection are invariant.

The aim of this paper is twofold: First we want to generalize the
notion of a weakly conserved energy current \`a la (1.2) to the
metric-affine
spacetime of MAG.  In Sec.4, we will derive such a current $\varepsilon
_{\rm \,MA}$, which generalizes the $\varepsilon\, _{\rm R}$ of (1.2).

Moreover, in Sec.5 we will relax the Killing constraints on the vector
field $\xi$ and consider {\it conformal} Killing vector fields in order
to obtain also conserved dilation and proper
conformal currents. In Minkowski spacetime, this construct is
well-known. Here we are generalizing it for the first time to MAG,
in the framework of which it finds its most natural embedding.

In Minkowski spacetime
(cf. Jackiw[3]) $${\cal D}^\alpha=x^\gamma\,\tilde t _\gamma{}^\a \,,
\eqno(\z)$$
$${\cal K}^{\a\beta}=[2x^\beta x^\gamma-g^{\beta\gamma}x^2]\, \tilde
t_\gamma{}^\alpha\eqno(\z)$$ are
the dilation current and the proper conformal
current, respectively, if $\tilde t^{\alpha\beta}$ represents the
improved energy-momentum tensor. From the divergence of these
currents we see
that both, scale and proper conformal invariance, are broken by
the trace $\tilde t_\mu{}^\mu$ of the improved energy tensor. Since
MAG is formulated in a locally dilation invariant way, we expect that
an appropriately constructed conformal current $\varepsilon \, _{\rm C}$
emerges in terms of a conformal Killing vector.
This expression is displayed in Eq.(6.10).  We will prove that it is
`weakly' conserved, provided the energy-momentum is tracefree.
Therefrom, in Sect.6, we will construct
an `improved' energy--momentum current
$\tilde\varepsilon \, _{\rm MA}$ which has a `soft' trace also for scalar
fields, as exemplified by the dilaton,
provided the mild constraint (6.11) holds.
All the previously discussed energy--momentum
expressions can be recovered as special cases of our new expressions
(5.10) or (6.10). We conclude our paper in Sec.7 with a relevant application
of the Ogievetsky theorem.

\bigskip\goodbreak
\sectio{\bf Metric-affine spacetime in brief}

The geometrical variables of such a metric-affine spacetime are the
forms $(g_{\alpha\beta},\vta^\alpha,$
$\Gamma_\alpha{}^\beta)$ with an appropriate transformation behavior
under the local $GL(n,R)$ group in $n$ dimensions. The components
$g_{\alpha\beta}$ of the metric $g$ are 0-forms, the coframe and the
connection are 1-forms. In MAG, the {\it nonmetricity}
$$Q_{\alpha\beta}:=-Dg_{\alpha\beta}\quad
\Rightarrow\quad Q^{\alpha\beta}=Dg^{\alpha\beta}\, ,\eqno(\z)$$
besides torsion $T^\alpha$ and curvature $R_\alpha{}^\beta$, enters
the spacetime arena as a new field strength. The gauge covariant derivative
$$D\Psi =d\Psi + \,^{\dagger}\!\Gamma_{\alpha}{}^{\beta}\wedge
\rho (L_{\beta}{}^{\alpha})\Psi + {w\over 2}Q\wedge\Psi\eqno(\z)$$
contains the volume--preserving $SL(n,R)$ connection
$ \,^{\dagger}\!\Gamma_{\alpha}{}^{\beta}$ and a volume--changing
piece depending on the Weyl 1-form $Q:=(1/n)\,Q^\alpha{}_\alpha $.
The matter fields are allowed to be densities, that is, under the
scale part $\Lambda_\a{}^\b=\delta_\a^\b\,\Omega$ of the $GL(n,R)$
gauge transformation the matter field will transform as $\Psi^\prime
=\Omega^w\,\Psi$. The weight $w$ depends on the $SL(n,R)$ index
structure and the density character.
Our notation is the same as in Refs.~[2].

%----------------------table beginning-------------------------
$$\vbox{\offinterlineskip
\hrule
\halign{&\vrule#&\strut\quad\hfil#\quad\hfil\cr
height2pt&\omit&&\omit&&\omit&\cr
& {\bf potential } && {\bf field strength } && {\bf Bianchi identity } &\cr
height2pt&\omit&&\omit&&\omit&\cr
\noalign{\hrule}
height2pt&\omit&&\omit&&\omit&\cr
& metric $g_{\alpha \beta }$ && $Q_{\alpha \beta }= -Dg_{\alpha \beta }$ &&
$DQ_{\alpha \beta }= 2 R_{(\alpha \beta )}$ &\cr
& coframe  $\vartheta ^\alpha $  &&
$T^\alpha = D\vartheta ^\alpha $ && $DT^\alpha = R_\mu {}^\alpha \wedge
\vartheta ^\mu $ &\cr
& connection $ \Gamma _\alpha {}^\beta $ && $ R_\alpha {}^\beta
= d \Gamma _\alpha {}^\beta + \Gamma _\mu {}^\beta \wedge
\Gamma _\alpha {}^\mu $ && $D  R_\alpha {}^\beta =0 $ &\cr
height2pt&\omit&&\omit&&\omit&\cr}
\hrule}$$
%----------------------table end-------------------------
\bigskip
Following Wallner [4], we use ${\cal L}_\xi$ to denote the usual Lie
derivative of a general geometric object and $\ell_\xi$ to denote the
restriction of the Lie derivative to differential forms, for which one
can show that $$\ell_\xi := \xi\rfloor\,d\ +\
d\,\xi\rfloor\,.\eqno(\z)$$ The ``gauge covariant Lie derivative'' of a
form is given by
$$\L:=\xi\rfloor\,D\ +\ D\,\xi\rfloor\,.\eqno(\z)$$
As an example, we obtain for the connection
$${\cal L}_\xi\Gamma_\a{}^\b=(\ell_\xi-\d_\xi)\,\Gamma_\a{}^\b
\qquad{\rm with}\qquad \d_\xi\Gamma_\a{}^\b=-D\,\bigl(e_\a\rfloor
(\ell_\xi\vta^\b)\bigr)\,.\eqno(\z)$$
Later on, we will employ the covariant exterior derivative
${\buildrel\frown\over{D}}$
with respect to the {\it transposed} connection
$${\buildrel\frown\over{\Gamma}}_{\alpha}{}^{\beta}:=
\Gamma_{\alpha}{}^{\beta} +e_{\alpha}\rfloor T^{\beta}\,.\eqno(\z)$$
Its somewhat unclear role becomes more transparent by the following
observation: If applied to the vector components $\xi^{\alpha}$, the
transposed derivative is identical to the gauge-covariant Lie
derivative of the coframe, i.e.
$${\L}_{\xi}\, \vartheta^{\alpha}
\equiv {\buildrel\frown\over{D}}\, \xi^{\alpha}\,. \eqno(\z)$$

\bigskip\goodbreak
\sectio{\bf Lagrange-Noether machinery}

The external currents of a matter field are those currents which are
related to {\it local} spacetime symmetries. On a fundamental level, we adopt
the view that fundamental matter is described in terms of
infinite-dimensional spinor or tensor representations of $SL(4, R)$,
the {\it manifields} $\Psi$, see Ne'eman [5].

In a first order formalism (cf. Nester [6] and Kopczy\'nski [7])
we assume that the material Lagrangian
$n$-form for these manifields depends most generally on $\Psi$,
$d\Psi$, and the potentials $g_{\alpha\beta}$, $\vartheta^{\alpha}$,
$\Gamma_{\alpha}{}^{\beta}$. According to the minimal coupling
prescription, derivatives of these potentials are not permitted.  We
usually adhere to this principle. However, Pauli type terms and the
Jordan-Brans-Dicke term $\bar{\Phi}\Phi
\,R^{\alpha\beta}\wedge\eta_{\alpha\beta}$ may occur in conformal
models with `improved' energy-momentum tensors or in the context of
symmetry breaking.  Therefore, we have developed [2] our Lagrangian
formalism in sufficient generality in order to cope with such models
by including in the Lagrangian also the derivatives
$dg_{\alpha\beta}$, $d\vartheta^{\alpha}$, and
$d\Gamma_{\alpha}{}^{\beta}$ of the gravitational potentials: $$L= L(
g_{\alpha\beta}\,, dg_{\alpha\beta}\, ,
\vartheta^\alpha\, , d\vartheta^\alpha\, ,\Gamma_{\alpha}{}^{\beta}\, ,
d\Gamma_{\alpha}{}^{\beta}\, ,\Psi, d\Psi)\,.\eqno(\z)$$
As a further bonus, we can then easily read off the Noether
identities for the gravitational gauge fields in $n > 2$ dimensions
(the restriction is related to the introduction of conformal symmetry
in Sec. 5).

For such a gauge-invariant Lagrangian $L$, the variational derivative
$\delta L/\delta\Psi$ becomes identical to the $GL(n,R)$-covariant
{\it variational derivative} of $L$ with respect to the $q$-form
$\Psi$: $${{\delta L}\over{\delta\Psi}} :=
{{\partial L}\over{\partial\Psi}} - (-1)^{q}D {{\partial
L}\over{\partial (D\Psi)}}\,. \eqno(\z)$$ The {\it matter currents}
are the metric stress-energy $\sigma^{\a\b}$, the cano\-nical
energy-momen\-tum $\Sigma_\a$, and the hypermomentum $\Delta^\a{}_\b$,
which is asymmetric in $\alpha$ and $\beta$.
They are given by
$$\sigma^{\alpha\beta} := 2{{\delta L}\over{\delta g_{\alpha\beta}}} =
2{{\partial L}\over{\partial g_{\alpha\beta}}}+ 2D{{\partial
L}\over{\partial Q_{\alpha\beta}}}\, , \eqno(\z)$$ $$\Sigma_{\alpha}:=
{{\delta L}\over{\delta\vartheta^{\alpha}}} = {{\partial
L}\over{\partial\vartheta^{\alpha}}}+ D {{\partial L}\over{\partial
T^{\alpha}}}\, ,\eqno(\z)$$ and $$\eqalign{\Delta^{\alpha}{}_{\beta}
:= {{\delta L}\over{\delta\Gamma_{\alpha}{}^{\beta}}} &=
L^{\alpha}{}_{\beta}\,\Psi\wedge{{\partial L}\over{\partial
(D\Psi)}}\cr &+ 2 g_{\beta\gamma}\,{{\partial L}\over{\partial
Q_{\alpha\gamma}}}+
\vartheta^{\alpha}\wedge
{{\partial L}\over{\partial T^{\beta}}} +
D {{\partial L}\over{\partial R_{\alpha}{}^{\beta}}}\, .\cr}
\eqno(\z)$$

The explicit form of the {\it canonical energy-momentum current} reads
$$\eqalign{\Sigma_\alpha &=
e_\alpha\rfloor L-(e_\alpha\rfloor D\Psi)\wedge
  {\partial L\over\partial D\Psi}-(e_\alpha\rfloor\Psi)\wedge
  {\partial L\over\partial\Psi}\cr
&\quad\cr
&\quad -(e_{\alpha}\rfloor Q_{\beta\gamma})\,
{\partial L\over{\partial Q_{\beta\gamma}}} -
(e_{\alpha}\rfloor T^\beta)\wedge
{\partial L\over\partial T^\beta}+
D{\partial L\over\partial T^\alpha}-
(e_{\alpha}\rfloor R_{\beta}{}^{\gamma})\wedge
{\partial L\over{\partial R_{\beta}{}^{\gamma}}}
\,.\cr }\eqno(\z)$$
The first line in (3.6) represents the result known in the context
of {\it special} relativistic classical field theory.
The last $\Psi$-dependent term in (3.6) vanishes for a $0$-form, as is
exemplified by a scalar or the Dirac field. The second line in (3.6) accounts
for possible Pauli terms and is absent in the case of  minimal coupling.

In the following, our Noether currents, aside from corresponding to the
gauge action of $GL(n,R)$, involve local translations as well. In
Minkowski spacetime, we could restrict the transformations to constant
parameters and apply $A(n,R)=R^n\semidirect GL(n,R)$ globally, still
obtaining the familiar conservation laws, including those of the dilations
and conformal transformations (1.3, 1.4). In our present paper, however,
we treat the whole $A(n,R)$ locally, even though one should observe that the
``gauging'' of $R^n$ transcends the definition of a Lie algebra and the
corresponding bundle Maurer-Cartan equations.

The {\it first Noether identity} in MAG takes the form
$$\eqalign{D\Sigma_\alpha
  \equiv&(e_\alpha\rfloor T^\beta)\wedge\Sigma_\beta+
  (e_\alpha\rfloor R_{\beta}{}^{\gamma})\wedge\Delta^{\beta}{}_{\gamma}
-{1\over2}(e_{\alpha}\rfloor
Q_{\beta\gamma})\,\sigma^{\beta\gamma} \cr
  &+(e_\alpha\rfloor D\Psi)\,{\delta L\over\delta\Psi}
  +(-1)^{p}(e_\alpha\rfloor\Psi)\wedge D{\delta L\over\delta\Psi}\cr
  \cong &(e_\alpha\rfloor T^\beta)\wedge\Sigma_\beta+
  (e_\alpha\rfloor R_{\beta}{}^{\gamma})\wedge\Delta^{\beta}{}_{\gamma}
-{1\over2}(e_{\alpha}\rfloor
Q_{\beta\gamma})\,\sigma^{\beta\gamma}. \qquad {\rm (1st)}\cr}
\eqno(\z)$$
Usually, our first result is given in the {\it strong} form, where no field
equation is invoked. The {\it weak} identity, which
is denoted by $\cong$, holds only
provided the matter field equation $\delta L/\delta\Psi=0$ is satisfied.

The {\it second Noether identity} in MAG reads
$$D\Delta^{\alpha}{}_{\beta} +
\vartheta^{\alpha}\wedge\Sigma_{\beta} -
g_{\beta\gamma}\>\sigma^{\alpha\gamma}\equiv -
L^{\alpha}{}_{\beta}\Psi\wedge{\delta L\over \delta\Psi}\cong 0\, .
\qquad {\rm (2nd)}\eqno(\z)$$

The dilational part of the {\it second} Noether identity can be easily
extracted directly from (3.8) by sheer contraction:
$$D\Delta + \vartheta^{\alpha}\wedge\Sigma_{\alpha}-
\sigma^{\alpha}{}_{\alpha}\equiv -
L^{\gamma}{}_{\gamma}\Psi\wedge{\delta L\over \delta\Psi}\cong 0\,
\,.\eqno(\z)$$
Only the trace piece $\vartheta^{\alpha}\wedge\Sigma_{\alpha}$
of the energy-momentum current
contributes to this dilational identity, cf. [8, 9].
This identity plays an important role for
the approximate scale invariance in the high-energy-limit of
particle physics (see [10]) and in the construction of the improved current.

\bigskip\goodbreak
\sectio{\bf Conserved currents in MAG}

In MAG, the connection is an independent field variable.
The corresponding matter current, coupled to it, will be the hypermomentum
$\Delta ^\alpha{}_\beta$. We require
$\xi=\xi^{\alpha}\, e_{\alpha}$ to be a Killing vector field for metric
{\it and} connection,

$$ {\cal L}_{\xi}\, g = \left(\L g_{\alpha\beta}+2g_{\gamma (\alpha}
e_{\beta )}\rfloor\L\vartheta ^\gamma \right)\; \vartheta ^\alpha\otimes
\vartheta^\beta =0\;,\qquad\quad
\qquad{\cal L}_{\xi}\Gamma_{\alpha}{}^{\beta} =0\, . \quad\eqno(\z)$$
According to (2.3), (2.5), and (2.7), these conditions can be recast
into the form
$$g_{\gamma (\alpha}\, e_{\beta )}\rfloor
{\buildrel\frown\over{D}}\, \xi^{\gamma} -
{1\over 2}\xi\rfloor Q_{\alpha\beta}=0\; ,
\qquad D(e_\a\rfloor {\buildrel\frown\over{D}}\, \xi^{\beta})
+\xi\rfloor R_{\alpha}{}^{\beta}=0\, . \eqno(\z)$$
Note that the first equation of (4.2) can be written alternatively, in terms
of the Riemannian derivative, as $e_{(\alpha}\rfloor {\buildrel {o}\over
{D}}\xi_{\beta )}=0$, compare the second equation of $(1.1)$.

Let us define the current $(n-1)$-form
$$\varepsilon\, {}_{\rm MA}:=\xi^{\alpha}\, \Sigma_{\alpha}
+(e_{\beta}\rfloor {\buildrel\frown\over{D}}\xi^{\gamma})\,
\Delta^{\beta}{}_{\gamma} \quad . \eqno(\z)$$
We compute its exterior covariant derivative, substitute
the two Noether identities (3.7) and (3.8), and reshuffle the emerging
expressions:
$$\eqalign{d\,\varepsilon_ {\rm \,MA} &= (D\xi^{\alpha})\wedge\Sigma_{
\alpha}+\xi^{\alpha}\, D\Sigma_{\alpha}
+D(e_{\beta}\rfloor {\buildrel\frown\over{D}}\xi^{\gamma})\wedge
\Delta^{\beta}{}_{\gamma}+(e_{\beta}\rfloor
{\buildrel\frown\over{D}}\xi^{\gamma})\, D\Delta^{\beta}{}_{\gamma} \cr
&\cong  (D\xi^{\alpha})\wedge\Sigma_{\alpha} +
(\xi\rfloor T^{\beta})\wedge\Sigma_\beta+
  (\xi\rfloor R_{\beta}{}^{\gamma})\wedge
\Delta^{\beta}{}_{\gamma}
-{1\over2}(\xi\rfloor  Q_{\beta\gamma})\,\sigma^{\beta\gamma} \cr
&\quad + D(e_{\beta}\rfloor {\buildrel\frown\over{D}}\xi^{\gamma})\,
\wedge\Delta^{\beta}{}_{\gamma}+(e_{\beta}\rfloor
{\buildrel\frown\over{D}}\xi^{\gamma})\,( \sigma^{\beta}{}_{\gamma} -
\vartheta^{\beta}\wedge\Sigma_{\gamma})\cr
&=\Bigl[ {\buildrel\frown\over{D}}\xi^{\alpha}-
\vartheta^{\beta}\,(e_{\beta}\rfloor
{\buildrel\frown\over{D}}\xi^{\alpha})\Bigr]\wedge \Sigma_{\alpha}
+\Bigl[ (De_{\beta}\rfloor {\buildrel\frown\over{D}}\xi^{\gamma} ) +
\xi\rfloor R_{\beta}{}^{\gamma}\Bigr]\wedge
\Delta^{\beta}{}_{\gamma}\cr
&\quad + \Bigl[g_{\alpha(\beta }\, e_{\gamma )}\rfloor
{\buildrel\frown\over{D}}\, \xi^{\alpha} -
{1\over 2}\xi\rfloor Q_{\beta\gamma}\Bigr]\sigma^{\beta\gamma}
\, .\cr }\eqno(\z)$$
While transforming the exterior derivative into the gauge covariant
derivative (2.2),
we assumed that $\varepsilon$ has zero weight $w_{\varepsilon}$, which
is in accordance with the weight zero for the Lagrangian and a usual vector
field $\xi $. Moreover,
we recognize that the first square bracket vanishes identically
because of the relation $p\psi = \vartheta^\alpha\wedge(e_\alpha
\rfloor\psi)$,
which is valid for any $p$-form.
In view of the generalized Killing equations (4.2), also the other
expressions in the square brackets vanish. Thus, the
current (4.3) is weakly conserved

$$ d\, \varepsilon\, _{\rm MA} \cong 0 \, .\eqno(\z)$$

For the Riemann-Cartan spacetime of the Einstein-Cartan-Sciama-Kibble
theory a similar result has been obtained by Trautman [11] and, for
the linearized case, by Tod [12]. The corresponding current reads
$$\varepsilon\, {}_{\rm RC}:=\xi^{\alpha}\, \Sigma_{\alpha}
+(e_{\beta}\rfloor {\buildrel\frown\over{D}}\xi^{\gamma})\,
\tau^{\beta}{}_{\gamma}\, ,\eqno(\z)$$
where the spin current is defined according to $\tau^{\b\c}:=
\Delta^{[\b\c]}$.
In Audretsch et al.\ [13], this current
was used to construct a Hamiltonian for the Dirac field.

Provided a timelike Killing vector field exists, we have obtained, via
(4.5), a globally
conserved energy $\int_S\varepsilon_{\rm \,MA}$. Our deduction of this
expression follows the pattern laid out in GR, but generalizes it to a
metric-affine spacetime.  Some steps of this deduction resemble
Penrose's recent {\it local mass construction} [14], except that we
refrain from using spinor or twistor methods at this stage.

\bigskip\goodbreak
\sectio{\bf Conserved dilation and proper conformal currents}

If the metric-affine spacetime admits a conformal symmetry,
an important generalization of (4.3) can be constructed as follows:
Let $\xi=\xi^{\alpha}\, e_{\alpha}$ be a conformal Killing
vector field such that the Lie derivative of the metric $g$ and the
connection $\Gamma_{\alpha}{}^{\beta}$
read\myfoot{Eq. $(5.1)_2$ implies the vanishing of
the trace free part of $(4.2)_2$. The same would hold, too, by requiring
$(5.1)_2$ for the volume--preserving connection instead.}
$$ {\cal L}_{\xi}\, g =
\omega \, g\,,\qquad\qquad{\cal L}_{\xi}%{}^\dagger
\Gamma_{\alpha}{}^{\beta} =
{1\over 2}\delta_\alpha^\beta\, d\omega\, . \eqno(\z)$$
The same algebra as in Sect.4  yields
$$g_{\c(\alpha}\, e_{\beta )}\rfloor
{\buildrel\frown\over{D}}\, \xi^{\gamma} -
{1\over 2}\xi\rfloor Q_{\alpha\beta}={1\over 2}g_{\alpha \beta }\,\omega\,,
 \qquad\qquad D(e_\a\rfloor {\buildrel\frown\over{D}}\, \xi^{\beta})
+\xi\rfloor R_{\alpha}{}^{\beta}={1\over 2} \delta_\alpha^\beta\, d\omega
\, , \eqno(\z)$$
compare with (4.2), which we recover for $\omega=0$. It follows from
(5.1)$_1$ that $\xi$ generates a transformation, parametrized by $\lambda$,
of the spacetime mainfold such that the metric undergoes the special [15]
conformal change $g\rightarrow \tilde g= e^{\lambda\omega}\, g$.
For a given geometry, the scalar function $\omega =\omega (x)$ is
determined by the trace of $(5.2)_1$, i.e. by
$$\omega = {2\over n}e_\gamma\rfloor{\buildrel\frown\over {D}}\xi^\gamma
-\xi\rfloor Q\, .\eqno(\z)$$
Thus, in metric-affine spacetime the conformal Killing equation for
the metric reads
$$ g_{\c(\alpha}\, e_{\beta )}\rfloor
{\buildrel\frown\over{D}}\, \xi^{\gamma} -
{1\over n}g_{\alpha\beta}\, e_\gamma\rfloor{\buildrel\frown\over{D}}
\xi^\gamma={1\over 2}\xi\rfloor Q\quer_{\alpha\beta}\, ,\eqno(\z)$$
where $ Q\quer_{\alpha\beta} :=Q_{\alpha\beta}-g_{\alpha\beta}\, Q$
is the tracefree part of the nonmetricity.

Again we compute the exterior derivative of (4.3), but now under the
assumption that the vector field $\xi $ is a conformal Killing field
obeying (5.1) or (5.2), respectively. Then the expressions in
the last two square
brackets in (4.4) do not vanish any more. Rather we find with the aid
of the Noether identity (3.9) the relation
$$d\,\varepsilon_ {\rm \,MA} \cong {1\over 2} \, D\omega \wedge \Delta ^\a{}_\a
+{1\over 2}\omega\,\sigma ^\alpha{}_\alpha\;
\cong  d\,({1\over 2}\omega\,\Delta )+{1\over 2}\omega\,(\vartheta^\alpha
\wedge\Sigma_\alpha)\, .  \eqno(\z)$$
In the case of vanishing $\omega $, we recover (4.5).
If $\omega$ does not vanish, eq.(5.5) yields
$$d\,(\varepsilon _{\rm \,MA}-{1\over 2} \omega\,\Delta)\cong
{1\over 2}\omega\,(\vartheta^\alpha\wedge
\Sigma_\alpha)\,.\eqno(\z)$$
For conformally invariant gauge theories, such as the Maxwell
or the Yang-Mills vacuum theory, the trace
$\vta^\a\wedge\Sigma_\a$ of the energy-momentum current vanishes and
(5.6) provides already the conserved quantity
$$\varepsilon\, _{\rm C}:=\varepsilon\,{}_{\rm MA}
-{1\over 2}\omega\,\Delta \, . \eqno(\z)$$

Since the traces of the energy-momentum current and the
hypermomentum current
$(n-1)$--forms are crucial here, we decompose these currents into
their tracefree pieces and their traces, respectively:
$$\Sigma_\alpha={\nearrow\!\!\!\!\!\!\!\!\Sigma}{}_\alpha +
{1\over n}\,e_\alpha\rfloor (\vartheta^\gamma\wedge\Sigma_\gamma )\, ,\qquad
\qquad\Delta^\alpha{}_\beta ={\nearrow\!\!\!\!\!\!\!\!\Delta}\,{}^\alpha{}
_\beta +{1\over n}\,\delta^\beta_\alpha\,\Delta\, .
\eqno(\z)$$
We substitute first (4.3) and then (5.8) into (5.7). This yields
$$\varepsilon\, _{\rm C}=\xi^\alpha{\nearrow\!\!\!\!\!\!\!\!\Sigma}
{}_\alpha +{1\over n}\,\xi\rfloor (\vartheta^\alpha\wedge\Sigma_\alpha )
+(e_\beta\rfloor{\buildrel\frown\over {D}}\xi^\gamma)\,
{\nearrow\!\!\!\!\!\!\!\!\Delta}\,{}^{\beta}{}_{\gamma}+
{1\over n}\left[(e_\alpha\rfloor{\buildrel\frown\over {D}}\xi^\alpha )
-{n\over 2}\omega \right]\,\Delta \, .\eqno(\z)$$
If we apply the trace (5.3) of one of the conformal Killing relations,
we obtain
$$\varepsilon\, _{\rm C}=\xi^\alpha{\nearrow\!\!\!\!\!\!\!\!\Sigma}{}
_\alpha +(e_\beta\rfloor{\buildrel\frown\over {D}}\xi^\gamma)\,
{\nearrow\!\!\!\!\!\!\!\!\Delta}\,{}^{\beta}{}_{\gamma}+
{1\over n}\,\xi\rfloor (\vartheta^\alpha\wedge\Sigma_\alpha )+
{1\over 2}(\xi\rfloor Q)\,\Delta\, . \eqno(\z)$$

Thus we have found generalizations of the well-known dilation
and proper conformal currents in Minkowski spacetime (cf. Ref. [3]) to
a metric-affine spacetime. Such a spacetime provides the most natural
gravitational background for these currents.

\bigskip\goodbreak
\sectio{\bf Improved currents with a `soft' energy--momentum trace for
scalar fields}
\bigskip

Following Isham et al.\ [16], we consider the
so-called {\it dilaton field} $\sigma(x)$ being immersed into our
MAG framework.

On the classical level, we can assume that the scalar field carries
canonical dimensions , i.e. $(length)^{-1}$ in $n=4$ dimensions.
With respect to a generic conformal change
$g\rightarrow \tilde g=\Omega (x) g$ of the underlying metric structure,
the scalar field will then transform according to
$$\sigma (x) \rightarrow \tilde \sigma (x) = \Omega (x)^{-(n-
2)/4}  {} \sigma (x) \, ,\eqno(\z) $$
i.e. as a scalar density of weight $w_\sigma = (2-n)/2$.
Then the gauge--covariant exterior derivative (2.2) is given by
$$ D \sigma := \left( d - {n-2 \over 4}\>
Q \right) \sigma \;  \eqno(\z) $$
which, due to the inhomogeneous transformation of the Weyl 1--form
under conformal changes, is conformally covariant.

The dilaton Lagrangian with a polynomial self--interaction
$n$--form $V(|\sigma |)$ is given by
$$L _{\sigma } =  L _{\square }  + V(|\sigma |),\quad
L _{\square } := {1\over 2}
\, D \sigma \wedge\, ^{*}\!  D  \sigma\; .  \eqno(\z) $$
Eq. (6.2) implies that its kinetic part is conformally invariant
in any dimensions.
By variation of (6.3) with respect to $\sigma$, we obtain the
scalar wave equation
$$ {\delta L _\sigma \over \delta\sigma} =
  \square\; \sigma +  {{\partial V(|\sigma |)}\over{\partial\sigma}} = 0, \;
\eqno(\z)$$
where
$$\square \sigma :=  -{ D}\, ^{*}\!
\left( D \sigma \right) \;
\eqno(\z) $$
is the d'Alembertian.

The trace of the dilaton's energy--momentum current, i.e.
$$\vartheta^{\alpha}\wedge\Sigma_{\alpha}(\sigma)  =
(n-2)\, L _{\square } + nV(|\sigma |)\eqno(\z) $$
does not vanish. Therefore, we need to ``improve'' $\Sigma_{\alpha}$ in
this respect. Since the kinetic part of the dilaton
Lagrangian (6.3) depends explicitly on the Weyl $1$--form $Q$,
the scalar field does also provide
an intrinsic dilation current.
According to (3.5), the latter is dynamically defined by
$$\Delta := \Delta ^\gamma {} _\gamma {} =
{{\delta L}\over {\delta \Gamma _{\gamma }{}^{\gamma}}} =
{2\over { n}}\> {{\delta L}\over{\delta Q}} =
{{2-n}\over{2n}}\> \sigma \,^{*}\!
D \sigma \; . \eqno(\z) $$

In our formalism we may define a ``new improved" energy--momentum current
for scalar fields by
$$\Theta_{\alpha} :=\Sigma_{\alpha} +
e_{\alpha}\rfloor D\Delta \, .
\eqno(\z) $$
After insertion of the field equation (6.4), we find for its trace the
``weak'' relation
$$\vartheta^{\alpha}\wedge\Theta_{\alpha}(\sigma)=
\vartheta^{\alpha}\wedge\Sigma_{\alpha}(\sigma)+ n D\Delta  \cong
nV(|\sigma |) - {{n-2}\over 2}\sigma {{\partial V(|\sigma |)}\over
{\partial\sigma}}.\eqno(\z) $$
Compared to (6.6), the kinetic term $L_ {\square }$ is absent in
(6.9).  Moreover, due to Euler's theorem for homogeneous functions,
the $\sigma^{2n/(n-2)}$ piece in the potential drops out. For a
polynomial potential $V(|\sigma|)$ of degree $p\leq 2n/(n-2)$, the
operator dimensionality is then smaller than $n$ (for $n\geq 4$).
Therefore, the new trace is indeed ``soft" in a momentum
representation in the sense of Jackiw ([3], p. 213; cf.  Kopczy\'nski
et al. [17]). Note that, for $n=4$, a pure $\sigma^{2n/(n-2)}$ model
is known to be renormalizable according to the criteria of power
counting.

The necessary modification of our globally conserved
current (5.10) is the following: We replace $\Sigma_{\alpha}$
by $\Theta_{\alpha}$ and define the
new `improved' current by
$$\eqalign{\tilde \varepsilon \,_{MA} &:=
\xi ^\alpha\Theta_\alpha +
(e_\beta\rfloor{\buildrel\frown\over {D}}\xi^\gamma)\,
{\nearrow\!\!\!\!\!\!\!\!\Delta}\,{}^{\beta}{}_{\gamma}+
{1\over 2}(\xi\rfloor Q)\Delta \cr
&=\xi ^\alpha\Sigma_\alpha +\xi\rfloor D\Delta+
(e_\beta\rfloor{\buildrel\frown\over {D}}\xi^\gamma)\,
{\nearrow\!\!\!\!\!\!\!\!\Delta}\,{}^{\beta}{}_{\gamma}+
{1\over 2}(\xi\rfloor Q)\Delta \cr
&=\varepsilon\, _{\rm C} +\xi\rfloor D\Delta\,.\cr }\eqno(\z) $$
Provided the scaling relation
$$ \ell_\xi(D\Delta)={n\over 2}\omega\, D\Delta \eqno(\z) $$
holds, we finally obtain
$$d \tilde \varepsilon \,_{MA}=
{1\over 2}\omega (\vartheta^\alpha\wedge\Theta_\alpha )\, .\eqno(\z)$$
This means that the divergence of the `new improved' current (6.10)
has indeed a soft trace also for scalar fields, as is required
quantum--theoretically. If $\vartheta^\alpha\wedge\Theta_\alpha$ is
vanishing, the scaling property (6.11) converts into one for the trace
of the canonical energy-momentum current.
In future we hope to find further physical
motivations for this hypothetical scaling property.

\bigskip\goodbreak
\sectio{\bf Emergence of the infinite-dimensional group of active
diffeomorphisms}
\bigskip
In (3.5) we defined the hypermomentum current in metric-affine spacetime.
In Minkowski spacetime we can expand the hypermomentum so as to exhibit
the hypersurface $(n-1)$-form $\eta_\alpha$. By using the improved
energy-momentum tensor, we obtain an orbital representation:
$$\tilde\Delta^\a{}_\b =\Delta^\a{}_\b{}^\gamma \eta_\gamma
\qquad , \qquad \Delta^\a{}_\b{}^\gamma =x^\a \tilde t _\b{}^\gamma
\, .\eqno(\z)$$
The antisymmetric piece $\Delta_{[\a \b ]}{}^{\c }$ is the angular momentum
tensor, whereas the shear components are given by the traceless part of
the symmetric piece $\Delta_{(\a \b)}{}^{\c }$. The integrated total
`charges' are given by (cf. the introduction)
$$D=\int x^\c\,\tilde t_\c{}^\a\eta_\a\qquad{\rm (dilation\>\> charge)}\,,
\eqno(\z)$$
$$K^\b=\int [2x^\b x^\c-g^{\b\c}\, x^2]\,\tilde t_\c{}^\a\eta_\a
\qquad{\rm (proper\>\> conformal\>\> charges)}\,,
\eqno(\z)$$
$$M_{[\b\c ]}=\int x_{[\b}\tilde t_{\c]}{}^\a\eta_\a
\qquad{\rm (angular\>\> momentum\>\> charges)}\,,
\eqno(\z)$$
$$S\!\!\!\!\!\!\nearrow _{(\b\c)}=\int [
x_{(\b}\tilde t_{\c)}{}^\a-{1\over n}
g_{\b\c}\,x^\delta \tilde t_{\delta}{}^\a]\eta_\a
\qquad{\rm (shear\>\> charges)}\,.
\eqno(\z)$$

Using Bohr's principle of correspondence, Ogievetsky [18] has shown
that the quantized system of shear $S\quer _{(\b\c)}$
and proper conformal $K^\b$ charges does not close algebraically and
generates the infinite algebra of diffeomorphism charges in $n$
dimensions:
$$^nL^{<p_0,...,p_{n-1}>\gamma\delta\epsilon\xi\eta\theta\cdots}_\a
=\int \, \prod_{i=0}^{n-1} (x_i)^{p_i} \;\tilde
t_\a{}^\b\eta_\b\,.\eqno(\z)$$ The algebraic relations are preserved
anholonomically. Thus, a metric-affine spacetime which admits a
conformal symmetry will have its frames locally invariant (in the {\it
active} operational sense of [19],[20]) under the group of analytical
diffeomorphisms.
This result overlaps with the fact that we have included in our affine
gauge approach local translations, i.e. active diffeomorphisms, except
that whereas the latter are only infinitesimal (their generators do not form
a Lie algebra anyhow), the Ogievetsky transformations can be integrated to
finite diffeomorphisms, provided we restrict to constant parameters and thus
do not require an infinite set of commutators.
The emergence of an explicit infinite Lie algebra may make it possible
to treat conformal fields in $n$-dimensions similarly to what is done
in the special case of two dimensions. In $n=2$, there is an
infinite--dimensional conformal algebra which is isomorphic to the algebra of
analytic two--dimensional diffeomorphisms [21]. In two dimensions this
feature constrains the fields and leads to the highly restrictive `fusion'
rules [21], which have recently put 2-dimensional conformal field theory
into the focus of interest of statistical mechanics and string theory.

The Ogievetsky algebra in $n$ dimensions is conceptually the
analog of the de Witt algebra in two dimensions and should possess a
quantum extension with central charges as in the Virasoro algebra. Neither
this extension nor the representation theory have been
investigated to date.
\bigskip
\centerline{\bf  Acknowledgment}
One of us (Y.N.) would like to thank Prof. M. Berger for
support and hospitality at the IHES.
\bigskip\goodbreak
\centerline {\bf References}

\newref
[1] R. Penrose and W. Rindler: {\it Spinors in spacetime}, Vol. 2.
(Cambridge University Press 1984); see also L.B. Szabados:
Canonical pseudotensors, Sparling's form and Noether currents,
preprint KFKI-1991-29/B, Hungarian Acad. Sci. Budapest (1991).
\newref
[2] F.W. Hehl, J.D. McCrea, E.W. Mielke, and Y. Ne'eman,
{\it Found. Phys.} {\bf 19}, 1075 (1989); {\it Phys. Rep.} (to be published);
J.D. McCrea, {\it Class. Quantum Grav.} {\bf 9}, 553 (1992).
\newref
[3] R. Jackiw, in: {\it Lectures on Current Algebra and its Applications},
B. Treiman, R. Jackiw, and D.J. Gross, eds. (Princeton University Press,
Princeton 1972) p.\ 97, see in particular p.\ 212.
\newref
[4] R. P. Wallner: ``Feldtheorie im Formenkalk\"ul'', Ph.D.-thesis,
University of Vienna 1982.
\newref
[5] Y. Ne'eman, in: {\it Spinors in physics and geometry}, Trieste, 11-13
Sept. 1986, A. Trautman and G. Furlan eds. (World Scientific, Singapore
1987), p. 313;
Y.\ Ne'eman, {\it Ann.\ Inst. Henri Poincar\'e} {\bf A28}, 369 (1978).
\newref
[6] J. M. Nester, in: {\it An Introduction to Kaluza-Klein Theories},
Workshop Chalk River/Deep River, Ontario 11-16 Aug. 1983,
H. C. Lee, eds. (World Scientific, Singapore 1984), p.83.
\newref
[7] W. Kopczy\'nski, {\it Ann. Phys.} (N.Y.), {\bf 203}, 308 (1990).
\newref
[8] F.W. Hehl and E.W. Mielke, {\it Wiss. Zeitschr. d.
Friedrich-Schiller-Universit\"at Jena} {\bf 39}, 58 (1990).
\newref
[9] I.M. Benn and W.P. Wood, {\it J. Math. Phys.} {\bf 33}, 2765 (1992).
\newref
[10] S. Coleman, {\it Phys. Rev.} {\bf 177}, 2239, 2247 (1969).
\newref
[11] A. Trautman, {\it Bull. Acad. Polon. Sci., S\'er. sciences math., astr.
et phys.} {\bf 20}, 895 (1972).
\newref
[12] K.P. Tod, in: {\it Twistors in Mathematics and Physics},
T.\ N.\ Bailey and R.\ J.\ Baston eds. (Cambridge University
Press, Cambridge 1990), p.\ 164.
\newref
[13] J. Audretsch, F.W. Hehl, and C. L\"ammerzahl in: {\it Proceedings
of the Bad Honnef School on Gravitation}, J. Ehlers and G. Sch\"afer,
eds. Lecture Notes in Physics, (Springer-Verlag, Berlin, to be published
1992).
\newref
[14] R. Penrose, {\it Proc. Roy. Soc.} (London)
{\bf A381}, 53 (1982); R. Penrose,
in: {\it Global Riemannian Geometry}, T.J. Willmore and
N.J. Hitchin, eds. (Ellis Harwood Limited, New York 1984), p. 203.
\newref
[15] E. W. Mielke, {\it Gen. Rel. Grav.} {\bf 8}, 321 (1977).
\newref
[16] C.J. Isham, Abdus Salam, and J. Strathdee, {\it Phys. Rev.} {\bf D2},
685 (1970).
\newref
[17] W. Kopczy\'nski, J.D. McCrea, and F.W. Hehl, {\it Phys. Lett.}
{\bf A128}, 313 (1988).
\newref
[18] V. I. Ogievetsky, {\it Lett. Nuovo Cimento } {\bf 8}, 988 (1973).
\newref
[19] F. Hehl, P. v. Heyde, G. D. Kerlick and J. M. Nester, {\it Rev. Mod.
Phys.} {\bf 48}, 393 (1976).
\newref
[20] Y.\ Ne'eman, in: {\it Differential Geometrical Methods in Mathematical
Physics}, K. Bleuler, H. R. Petry, and A. Reetz, eds. Lecture Notes in
Mathematics, Vol. 676 (Springer-Verlag, Berlin 1978), p.189.
\newref
[21] M. Kaku, {\it Strings, Conformal Fields, and Topology},
(Springer-Verlag, New-York 1991).
\bigskip\centerline{---------------}
\vfill\bye